\documentclass[10pt]{article}
\usepackage{float}
\usepackage[symbol]{footmisc} 
\usepackage[superscript,sort]{cite}
\usepackage{array}
\usepackage{bm}
\usepackage[pdftex,colorlinks=true,linkcolor=black,citecolor=black,urlcolor=black,bookmarks=true,breaklinks=true]{hyperref}
\newcolumntype{C}[1]{>{\centering\arraybackslash}p{#1}}
\usepackage[usenames]{color}
\usepackage{amsmath,amssymb}

\usepackage{amsthm,amscd,amsxtra,amsfonts,amsmath,amssymb,multirow}
\usepackage{wrapfig}
\usepackage[footnotesize]{caption}
\usepackage{subcaption}
\usepackage[tiny,compact]{titlesec}
\usepackage{threeparttable} %add footnote underneath the table
\usepackage{helvet}

\usepackage{booktabs}
\usepackage[table]{xcolor}
\usepackage{xcolor,colortbl}
\usepackage{tikz}
\usetikzlibrary{shapes,arrows}
\usepackage[T1]{fontenc}
\usepackage[linesnumbered,ruled]{algorithm2e} %algorithm
\titlespacing*{\section}{0pt}{*0}{*0}
\titlespacing*{\subsection}{0pt}{*0}{*0}
\titlespacing*{\subsubsection}{0pt}{*0}{*0} 
\titlespacing{\paragraph}{0pt}{*0}{*1}

\definecolor{MyPurple}{rgb}{1,0,1}

\newcommand{\beq}[1]{\begin{equation} \label{#1}}
\newcommand{\eeq}{\end{equation}}
\newcommand{\barray}{\begin{array}{ll}}
\newcommand{\earray}{\end{array}}

\begin{document}
\pagenumbering{roman}

\clearpage \pagebreak \setcounter{page}{1}
\renewcommand{\thepage}{{\arabic{page}}}

\title{TopP-S: Persistent homology based multi-task deep neural networks for simultaneous predictions of partition coefficient and aqueous solubility}

\author{Kedi Wu$^{1}$, Zhixiong Zhao$^{2}$, Renxiao Wang$^{3}$,  Guo-Wei Wei$^{1,4,5}$
\footnote{
Address correspondences  to Guo-Wei Wei. E-mail: wei@math.msu.edu}  \\
$^1$Department of Mathematics \\
Michigan State University, MI 48824, USA\\
$^2$School of Medicine \\
Foshan University, Foshan, Guangdong, 528000, China \\
$^3$State Key Laboratory of Bioorganic Chemistry, Shanghai Institute of Organic Chemistry, \\
Chinese Academy of Sciences, Shanghai, 200032, P.R. China \\
$^4$Department of Electrical and Computer Engineering \\
Michigan State University, MI 48824, USA \\
$^5$Department of Biochemistry and Molecular Biology\\
Michigan State University, MI 48824, USA
}

\date{\today}
\maketitle

\begin{abstract} 
Aqueous solubility and partition coefficient are important physical properties of small molecules.  Accurate theoretical prediction of aqueous solubility and partition coefficient plays an important role in drug design and discovery. The prediction accuracy depends crucially on molecular descriptors which are typically derived from theoretical understanding of the chemistry and physics of small molecules.  The present work introduces  an algebraic topology based method, called element specific persistent homology (ESPH), as a new representation of small molecules that is entirely different from  conventional chemical and/or physical representations. ESPH describes molecular properties in terms of multiscale and multicomponent topological invariants. Such topological representation is systematical,  comprehensive, and scalable with respect to  molecular size and composition variations. However, it cannot be literally translated into a physical interpretation. Fortunately, it is readily suitable for machine learning methods, rendering topological learning algorithms.  Due to the inherent correlation between solubility and partition coefficient, a uniform ESPH representation is developed for both properties, which facilitates  multi-task deep neural networks for their simultaneous predictions. This strategy leads to more accurate prediction of relatively small data sets. A total of six data sets is considered in the present work to validate the proposed topological and multi-task deep learning approaches. It is demonstrate that the proposed approaches achieve some of the most accurate predictions of aqueous solubility and partition coefficient.  Our software is available online at {\url{http://weilab.math.msu.edu/TopP-S/}}.

\end{abstract}

\maketitle

\newpage
{\setcounter{tocdepth}{5} \tableofcontents}
\clearpage \pagebreak \setcounter{page}{1}

\section{Introduction}\label{sec:Intro}

The partition coefficient, denoted  $P$ and defined to be the ratio of concentrations of a solute in a mixture of two immiscible solvents at equilibrium, is of great importance in pharmacology. It measures the drug-likeness of a compound as well as its hydrophobic effect on human body.  The logarithm of this coefficient, i.e., $\log P$, has proved to be one of  key parameters in drug design and discovery.  Optimal ${\log P}$ along with low molecular weight and low polar surface area plays an important role in governing kinetic and dynamic aspects of drug action. In particular, Hansch et al. \cite{leo:1995} gave a detailed description of how lipophilicity impacted pharmacodynamics. This being said, surveys show that approximately half of the drug candidates fail to reach market due to unsatisfactory pharmacokinetic properties or toxicity \cite{van:2003}, which indeed makes  $\log P$ predictions even more important.  

The extent of existing reliable experimental $\log P$ data is negligible compared to   tremendous compounds whose $\log P$ data are practically needed. Therefore, computational prediction of partition coefficient is an indispensable approach in modern drug design and discovery. Since the pioneering work of Hansch {\it et al.} \cite{hansch:1964,fujita:1964,leo:1971}, a large variety of octanol-water partition coefficient predictors has been developed over the past few decades. Many methods are generally called as quantitative structure-activity relationship (QSAR) models.
In general, these models can be categorized into atom-based additive methods, fragment/compound-based methods, and property based methods. One of atom-based additive methods, which was first proposed by Crippen and co-workers \cite{ghose:1987}, is essentially purely additive and effectively a table look-up per atom. Later on,  XLOGP3, a refined version of atom-based additive methods, was developed  \cite{chengtj:2007}. This approach considers various atom types, contributions from neighbors, as well as correction factors which help overcome known difficulties in purely atomistic additive methods. However additivity may fail in some cases, where unexpected contributions to $\log P$ occur, especially for complicated structures.   Fragment/compound based predictors, instead of employing information from single atom, are built at compounds or fragments level. Compounds or fragments are then added up with correction factors. Popular fragment  methods include KOWWIN \cite{meylan:1995,meylan:2000},  CLOGP \cite{leo:1993, chou:1979}, ACD/LOGP \cite{petrauskas:2000,walker:2004 }, KLOGP\cite{zhuhao:2005,sedykh:2006}. A major challenge for fragment/compound based methods is the optimal classification of ``building blocks". The number of fragments and corrections involved in current methods range from hundreds to thousands, which could be even larger if remote atoms are also taken into account. This fact may lead to technical problems in practice and may also cause overfitting in modeling. The third category is property-based. Basically property-based methods determine partition coefficient using properties, empirical approaches, three dimensional (3D) structures (e.g., implicit solvent models, molecule dynamics (MD) methods), and topological or electrostatic indices. Most of these methods are  modeled using statistical tools such as associative neural network (ALOGPS) \cite{tetko:2002,tetko:2004}.  It is worthy to mention that property-based methods are relatively computationally expensive, and depend largely on the choice of descriptors and accuracy of computations. This to some extent results in a preference of methods in the first two categories over those in the third.  
 
Another closely related chemical property is aqueous solubility, denoted by $S$, or its logarithm value $\log S$. In drug discovery and other related pharmaceutical fields,  it is of great significance to identify molecules with undesirable water solubility on early stages as solubility affects absorption, distribution, metabolism, and elimination processes (ADME) \cite{lipinski:1997,di:2006}.  QSPR models, along with atom/group additive models \cite{duan2012empirical, hou:2004,wang2009aqueous,wang2007development}, have been developed to predict solubility. For example, QSPR models assume that aqueous solubility correlates with experimental properties such as aforementioned partition coefficient and melting point \cite{yalkowsky:1980}, or molecular descriptors such as solvent accessible area. However, due to the difficulty of experimentally measuring solubility for certain compounds, the experimental data can contain errors up to 1.5 log units\cite{dearden:2006,dannenfelser:1991} and no less than 0.6 log units \cite{jorgensen:2002}. Such a high variability brings challenge to solubility prediction. 

Both partition coefficient and aqueous solubility reveal how a solute dissolves in solvent(s). Therefore it is reasonable to assume that there exists a shared feature representation across these two related tasks. In machine learning theory, multitask  (MT) learning is designed to take the advantage of  shared feature representations of correlated properties. It learns the so-called ``inductive bias'' from related tasks to improve accuracy using the same representation \cite{caruana:1998}. In other words, MT learning aims at learning a shared and generalized feature representation from multiple tasks and has brought new insights into the study of bioinformatics. Successful applications include splice-site and MHC-I binding prediction \cite{widmer:2012} in sequence biology, gene expression analysis, and system biology \cite{xuq:2011}. MT learning becomes more efficient when it is incorporated with  deep learning (DL) strategies. DL has successfully achieved state-of-the-art results in signal and information processing fields, such as speech recognition \cite{dahl:2012, deng:2013} and natural language processing \cite{socher:2012, sutskever:2014}, as well as toxicity prediction \cite{dahl:2014, majinshui:2015,mayr:2016,ramsundar:2015,KDWu:2017a} and aqueous solubility prediction \cite{lusci:2013}. 

Geometric descriptors are commonly used in machine learning to represent small molecules. In fact,  geometric representation of molecules, particularly macromolecules, often involves  too much structural detail and thus may become intractable for large and complex biomolecular data sets. In contrast, topology offers the highest level of abstraction  and truly metric free  representations of molecules. However, traditional topology incurs too much geometric reduction to be practically useful for molecules. Persistent homology bridges classical geometry and topology,  offering a multiscale  representation of molecular systems \cite{Edelsbrunner:2002,Zomorodian:2005}.  In doing so, it creates a family of  topologies via a filtration parameter, which leads to a one-dimensional topological invariants, i.e., barcodes of Betti numbers. The physical interpretations of Betti-0, Betti-1 and Betti-2 barcodes  are    isolated components, circles, and  cavities, respectively. Persistent homology has been applied to   the modeling and prediction of  nano particles, proteins and other biomolecules \cite{KLXia:2014c, KLXia:2015a, KLXia:2015d,KLXia:2015e, BaoWang:2016a}. Nonetheless, it was found that primitive persistent homology has very limited predictive power in machine learning based classification  of biomolecules \cite{ZXCang:2015}, which motivate us to introduce  element specific persistent homology (ESPH) to retain crucial  biological information during the topological simplification of geometric complexity   \cite{ZXCang:2017a, ZXCang:2017b, ZXCang:2017c}. ESPH has found it success in the predictions of  protein-ligand binding affinities  \cite{ZXCang:2017b, ZXCang:2017c} and mutation induced protein stability changes  \cite{ZXCang:2017a,  ZXCang:2017c}. Unfortunately, the representational and predictive power of persistent homology and ESPH for small molecules is essentially unknown.  Unlike proteins, small molecules involve a wide range of chemical elements and their properties are very sensitive to their chemical constitutions, symmetry and stereochemistry. Therefore, it is not clear whether persistent homology and ESPH are suitable descriptors for small  molecules. 

The  objective for this work is to explore the representationability and predictive power of ESPH for small molecules. To this end, we focus on  the analysis and prediction of small molecular solubility and partition coefficient.  Due to their relevance to drug design and discovery, relatively large data sets have been collected in the literature for these problems, which give rise to good data sets for validation of topological descriptors. To overcome the difficulty of small available data sets for certain problems, we construct topological learning by  integrating ESPH   and multitask deep learning  for partition coefficient and aqueous solubility predictions.  We show that ESPH provides a competitive description of relatively small drug-like molecules. Additionally, the inherent correlation between partition coefficient and aqueous solubility makes the multi-task strategy a viable approach in joint $\log P$ and $\log S$ predictions. 

The rest of this paper is structured as follow. In Section \ref{sec:methods}, we give an introduction to   element specific persistent homology and the construction of element specific topological descriptor (ESTD. The underlying motivation for ESTD is discussed and a concrete example is also presented in Section \ref{sec:estd}. In Section \ref{sec:algorithms}, we provide an overview of classic ensemble methods, multitask learning (MT) and deep neural network (DNN). The MT-DNN architecture is carefully formulated and illustrated in Section \ref{sec:mt-dnn}. 
In Section \ref{sec:results}, we first give an overview of data sets. The predictions of MT-DNN models as well as other  methods for partition coefficients and aqueous solubility are presented. Finally we wrap up the paper with some discussions in Section \ref{sec:discussion}.

\section{Datasets and Methods} \label{sec:methods}
This section is devoted to datasets, topological methods, machine learning algorithms and evaluation metrics. 

\subsection{An overview of data sets} \label{sec:data}
The primary work of this paper is to explore the proposed topology based multi-task methods for learning partition coefficient and aqueous solubility simultaneously. Data sets can naturally be divided into two parts -- one for partition coefficient prediction and the other for aqueous solubility prediction. Note that for partition coefficient prediction there are multiple test sets while the training set remains the same. 

\paragraph{Partition coefficient data sets} The training set used for partition coefficient prediction was originally compiled by {\it Cheng et al.} \cite{chengtj:2007} and consists of 8199 compounds, which is based on Hansch {\it et al.}'s compilation \cite{hansch:1995}. These compounds are considered to have reliable experimental $\log P$ values by Hansch (marked with * or $\checkmark$). In addition, three sets were chosen as test sets. The first test set, which is completely independent from the training set, contains 406 small-molecule organic drugs approved by the Food and Drug Administration (FDA) of the United States and represents a variety of organic compounds of pharmaceutical interests. This set was also compiled by {\it Cheng et al.} \cite{chengtj:2007}. The remaining two test sets, Star set and Non-star set, were publicly available and originated from a monograph of Avdeef \cite{avdeef:2012}. Star set comprises 223 compounds that are part of BioByte Star set and have been widely used to develop $\log P$ prediction method. The Non-star set contains 43 compounds that represent relatively new chemical structures and properties. The compound list and corresponding partition coefficient is available for download at \url{http://ochem.eu/article/17434}. 
We also made an attempt to expand our training set by searching the NIH database as other software packages use a large number of molecules for supervised learning. In this way, more than 3000 additional molecules were added to the training set.

\paragraph{Aqueous solubility data sets} In order to develop and validate prediction models for aqueous solubility, several well-defined aqueous solubility datasets were used. Firstly, a diverse data set of 1708 molecules proposed by Wang {\it et al.} \cite{wangjm:2007} was used to verify the predictive power of  descriptors. Both leave-one-out and 10-fold cross-validation were carried out on this set.  Furthermore, we also tested our models on a relatively small set with independent test sets \cite{hou:2004}. As Hou \cite{hou:2004} suggested, we also removed some molecules from the training set to ensure that training set and test set have no overlapping molecules.

\paragraph{Statistics of datasets} A summary of data sets used for the proposed models is given  in Table \ref{tab:datastats}.

\begin{table*}[!ht]
\centering
\caption{Summary of logP and logS data sets used}
\begin{tabular}{|c|c|p{0.1cm}|c|c|}
\hline
logP data & Number of molecules & & logS data & Number of molecules \\ \hline
logP train set & 8199  & &  logS train set 1 \cite{wangjm:2007} & 1708 \\
FDA  test set & 406 & & logS train set 2 \cite{hou:2004} & 1290 (1207 for test set 2)\\
Star test set & 223  & & logS test set 1 \cite{hou:2004} & 21 \\
Nonstar test set & 43 & & logS test set 1 \cite{hou:2004} & 120\\ \hline
\end{tabular}
\label{tab:datastats}
\end{table*}

\subsection{Element specific topological descriptors (ESTD) } {\label{sec:estd}}
A brief introduction is given to persistent homology and element specific persistent homology (ESPH), followed by a detailed example to illustrate the persistent homology characterization of small molecules. A refined version of ESPH and corresponding ESTD construction are also discussed.

\subsubsection{Persistent homology} 
Persistent homology is a branch of algebraic topology that defines topological spaces in terms of algebraic structures. It is a main workhorse for topological data analysis, which offers topological simplification  of data complexity. Unlike conventional physical or chemical approaches, persistent homology captures the underlying topological connectivity  of small molecules directly from atomic coordinates, i.e.,  point cloud data in $\mathbb{R}^n$. Mathematically, isolated atoms of a molecule are 0-simplices. The connectivity among atoms defines high dimensional simplexes. For example, linked two atoms (a line segment) gives rise to a 1-simplex and mutually linked 3 atoms in a triangular shape is called a 2-simplex.  A mutually linked four-atom  tetrahedron is  3-simplex and so on and so forth.  
 
Appropriate collection of  simplices forms a simplicial complex, which is a  topological space consisting of vertices (points), edges (line segments), triangles, and their high dimensional counterparts. Simplicial homology can be defined on the basis of simplicial complex and can be used to analyze topological invariants, i.e.,  Betti numbers. Physically, Betti-0, Betti-1 and Betti-2  describe numbers of independent components, rings and cavities, respectively.  However, it is important to note that topological connectivity  among a set of atoms in a molecule does not follow their physical relations, i.e., covalent bonds, hydrogen bonds and van der Waals bonds. Instead, it is defined by a filtration parameter, or artificial ball radius (not the atomic radius) for each atom. Therefore, at a given filtration radius, one obtains a set of simplices, and thus Betti numbers for a molecule. In persistent homology, the filtration radius varies continuously from zero to a large number until  no meaningful topological structure can be created further \cite{edelsbrunner:2000,Zomorodian:2005}. Therefore, persistent homology computes topological invariants of a given molecule at different spatial scales which correspond to different geometric shapes (simplices) and thus different topological connectivities. 
The persistence of topological invariants over the filtration for a given molecule can be recorded in barcodes or persistent diagrams \cite{CZOG05,Ghrist:2008}.  The barcode representation  of persistent homology is utilized in the present work to construct ESTDs.  
Readers are referred to Ref. \cite{KLXia:2014c} for   a detailed while still simple introduction of persistent homology.

\begin{figure}[ht!]
\small
\centering
\begin{subfigure}{0.49\textwidth}
\centering
\includegraphics[width=0.8\textwidth]{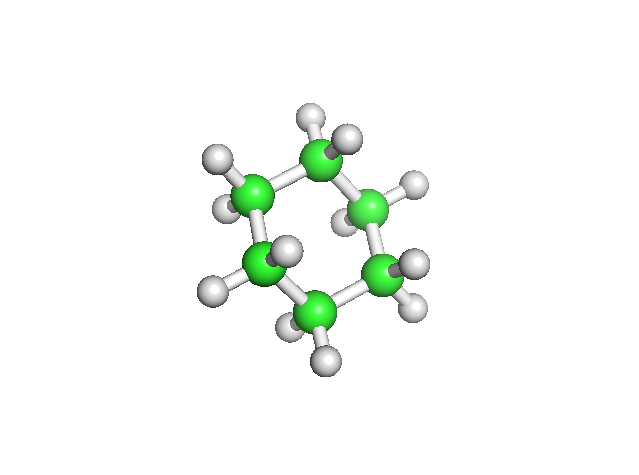}
\caption{All-element representation of cyclohexane}
\label{fig:cyc}
\end{subfigure}
\centering
\begin{subfigure}{0.49\textwidth}
\centering
\includegraphics[width=0.8\textwidth]{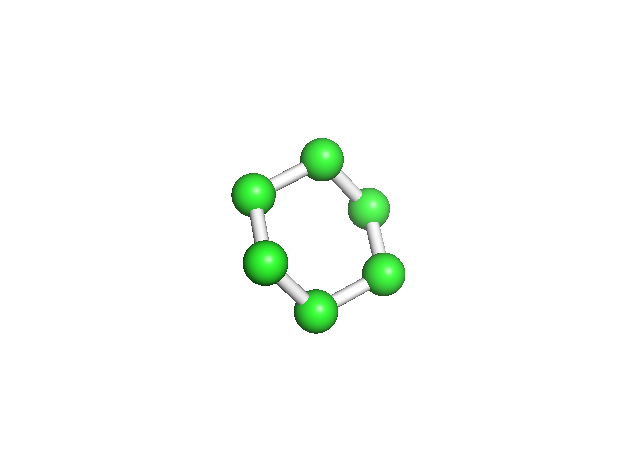}
\caption{C element representation of cyclohexane}
\label{fig:cyc-cc}
\end{subfigure}
\centering
\begin{subfigure}{0.49\textwidth}
\centering
\includegraphics[width=0.8\textwidth]{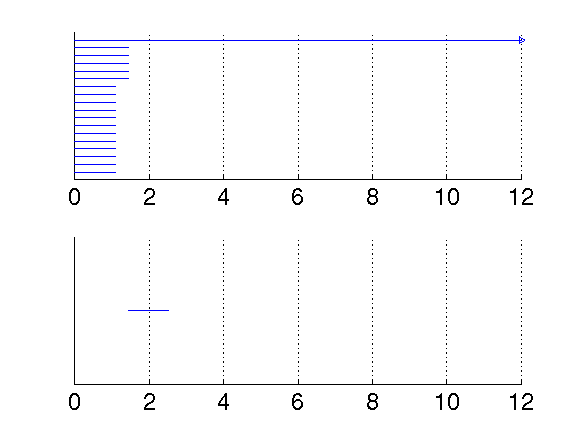}
\caption{Barcode plot of the all-element representation}
\label{fig:cyc-bar}
\end{subfigure}
\centering
\begin{subfigure}{0.49\textwidth}
\centering
\includegraphics[width=0.8\textwidth]{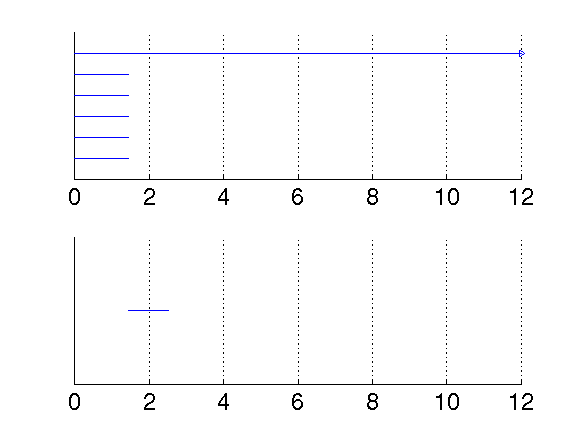}
\caption{Barcodes of the C element representation}
\label{fig:cyc-cc-bar}
\end{subfigure}
\caption{Cyclohexane and its persistent homology barcode plots. In subfigure(a) and (b), cyclohexane is shown with all elements and carbon element selected, respectively. In subfigure (c) and (d), from top to bottom, the results are for Betti-0 and Betti-1, respectively}
\label{fig:cyc-example}
\end{figure}

\paragraph{The necessity of ESPH and an example} 
Primitive persistent homology treats all atoms in an equal footing, which neglects chemical and physical properties of molecules. To obtain an accurate representation of a given molecule, it is necessary to at least distinguish different element types and construct element specific topological descriptors (ESTDs)  \cite{ZXCang:2017a, ZXCang:2017b, ZXCang:2017b}.  Figure \ref{fig:cyc-example} is a detailed example of how our ESTDs are calculated and how they can reveal the structure information of cyclohexane. An all-element representation of cyclohexan is given in Fig \ref{fig:cyc}, where carbon atoms are in green and hydrogen atoms are in white. As we can see from its barcode plot Fig. \ref{fig:cyc-bar}, there are 18 Betti-0 bars that correspond to 18 atoms at the very beginning, 12 of which disappear when the filtration value gets 1.08\AA. It indicates that each carbon atom has merged with its closest 2 hydrogen atoms as the filtration value becomes larger than the length of C-H bond and these three atoms are regarded as a single connected component. When the filtration value increases to 1.44\AA, a Betti-1 bar emerges which means that a hexagonal carbon ring is captured and there is only one connected component left. As the filtration value eventually exceeds the radius of the hexagon, the ring structure disappears and that is why there is no Betti-1 bar. The longest Betti-0 bar corresponds to the existence of the connected component. When only carbon atoms are selected (C element),  it is relatively straightforward to interpret the barcode plot. The cutoff where 5 Betti-0 bars disappear corresponds to the C-C bond length and the Betti-1 bar represents the existence of the hexagonal carbon ring. 

\paragraph{The challenge for primitive persistent homology}
Aforementioned persistent homology, although is able to capture the information such as covalent bonds between different atom types easily as shown in  Fig. \ref{fig:cyc-example}, does not necessarily reflect intramolecular interactions such as hydrogen bonds and van der Waals interaction, which is not ideal for the purpose of small molecule modeling. In other words, the Betti-0 bar between two atoms with certain hydrogen bonding or van der Waals cannot be captured since there already exist shorter Betti-0 bars between them (essentially covalent bonds). Thus it is important to redefine the distance between atom $i$ at $(x_i, y_i, z_i)$ and atom $j$ at $(x_j, y_j, z_j)$ as following:
\begin{equation}
    M_{i, j} = 
	\begin{cases}
    		d_{i,j}, & \text{if \hspace{0.1in}} d_{i,j} \ge r_i + r_j +|\Delta d |\\
    	    d_\infty,              & \text{otherwise}
	\end{cases}
	\label{eqn:newdistance}
\end{equation}
where $r_i$ and $r_j$ are the atomic radius of atoms $i$ and $j$, respectively, and $\Delta d$ is the bond length deviation in the data set. Here $d_\infty$ is a large number which is set to be greater than the maximal filtration value, and $d_{i,j}$ is the Euclidean distance between atom $i$ and atom $j$, or equivalently, 
\begin{equation}
d_{i,j}= \sqrt{(x_i-x_j)^2+(y_i-y_j)^2+(z_i-z_j)^2}. 
\label{eqn:distance}
\end{equation}
By setting the distance between two close atoms to an sufficiently large number, we should be able to capture intramolecular interactions since a connection longer than filtration value is automatically neglected during persistent homology computation.

\subsubsection{ESTD construction} 
Inspired by classic atom-additive models for partition coefficient prediction, we utilize a total of 61 basic element types calculated by antechamber \cite{wangjm:2004,wangjm:2006} using general amber force field (GAFF) \cite{wangjm:2004}. Atoms of given atom type and their appropriate combinations are selected to construct Vietoris-Rips complex and ESTDs are subsequently calculated. 

It is also important to construct ESTDs via a small bin size. As the example above shows, barcodes at different cutoffs give rise to a variety of information such as covalent and non-covalent bonds. Specifically we can divide the barcodes into several small bins and then extract features from each bin. A  complete list of ESTDs used in this study is shown in Ref.  \ref{tab:features}. Group 1 ESTDs focuses on different atom types, Group 2 is to capture the occurrences of non-covalent bonding and Group 3 mainly highlights the strength of non-covalent bonding and van der Waals interactions. Note that statistics of birth or death values in Group 3 refer to maximum, minimum, mean, and summation.   

\begin{table*}[!ht]
\centering
\caption{Detailed ESTD descriptions}
\begin{tabular}{c|c|c}
\hline
Feature group & Element used & Descriptors\\ \hline 
\multirow{3}{*}{Group 1}  & One element: $e$ & Counts of Betti-0 bars for each element type with a total \\
& where $e \in \{\mathrm{GAFF_{61}}\}$ & of 61 different element types calculated with GAFF \cite{wangjm:2004}\\ \hline 
\multirow{3}{*}{Group 2}  & Two element types: \{$a_i, b_j$\}, where & Counts of Betti-0 bars with birth or death values  \\ 
&  $a_i, b_j \in $ \{C, O, N\}, and $a_i \ne b_j$  & falling within each bin $B_i = [0.5i-0.5, 0.5i], i=1, \ldots, 10$ \\ \hline 
\multirow{3}{*}{Group 3}  & Two element types: \{$a_i, b_j$\}, where & Statistics of birth or death values \\ 
&  $a_i, b_j \in $ \{C, O, N\}, and $a_i \ne b_j$  &  for all Betti-0 bars (consider all birth and death values) \\ \hline 
\end{tabular}
\label{tab:features}
\end{table*} 

The essence of ESTDs is to offer new insight to small molecule modeling by topological modeling and topological learning. By constructing a topological feature vector $\mathbf{x_i^t}$ for the $i$-th molecule of task $t$, we are readily to combine topological learning with advanced machine learning algorithms, with further details to be discussed in Section \ref{sec:algorithms}. 

\subsection{Additional molecular descriptors}
In addition to ESTDs introduced in the previous subsection, we generate 633 2D molecule descriptors by CheomPy \cite{cao:2013} for each molecule. The feature pool contains feature groups such as  E-state descriptors. There are a total of 13 categories for these 2D molecule descriptors - 30 molecular constitutional descriptors,  35 topological descriptors, 44 molecular connectivity indices, 7 Kappa shape descriptors, 64 Burden descriptors, 245 E-state indices, 21 Basak information indices, 96 autocorrelation descriptors, 6 molecular property descriptors, 25 charge descriptors, and 60 MOE-type descriptors. A more detailed description of features and the CheomPy software are available on line at \url{https://code.google.com/archive/p/pychem/downloads}. 

In order to improve the overall performance, we also combine these features with ESTDs to create ESTD$^+$.  For consistency reasons, only molecules whose features can be calculated by both our ESTD software and ChemoPy software are used for training purpose. It is worth to mention that our ESTD approaches are applicable to all molecules whereas ChemoPy has difficulty in dealing with some molecules.  Separate results and discussions for different sets of descriptors will also be conducted in later sections. 

\subsection{Topological learning algorithms} \label{sec:algorithms}

In this section, we present methods and algorithms of topology based multi-task learning for simultaneous predictions of partition coefficient and aqueous solubility. Ensemble methods, including random forest (RF) and gradient boosting decision tree (GBDT), and neural network architectures are discussed. A detailed description of multi-task neural network is also provided. 

\subsubsection{Ensemble methods} Ensemble methods have been widely used to solve QSAR problems and have achieved some state-of-the-art results. They naturally handle correlations between descriptors and are generally insensitive to parametrization and feature selection due to bagging  operations. In this work we choose gradient boosting tree as our baseline method and implement it using the scikit-learn package \cite{scikit-learn} (version 0.13.1). The number of estimators for random forest is set to 8000 because a further increase in  the number does not essentially improve accuracy. For each set, 50 runs were done and the average of  50 predictions is taken as the final prediction. 

\begin{figure}[!ht]
\small
\centering
\includegraphics[scale=0.8]{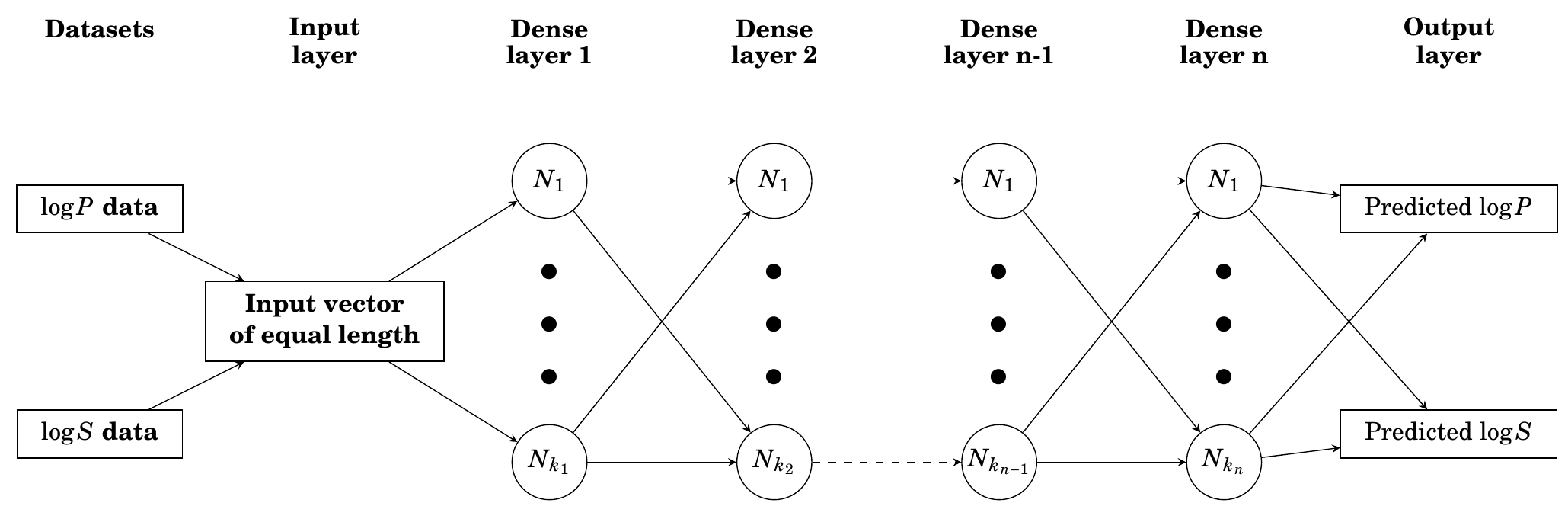}
\caption{An illustration of the MT-DNN architecture.}
\label{fig:MTDNN_arch}
\end{figure}

\subsubsection{Multi-task learning and deep neural networks} \label{sec:mt-dnn}
\paragraph{Multi-task learning} The idea of multi-task learning (MTL) is to learn `inductive bias' from related tasks to improve accuracy, using the same representation. In other words, MTL aims at learning a shared, generalized feature representation for multiple tasks and potentially gives better predictions. In our work, we assume that the underlying molecular mechanism of partition coefficient and aqueous solubility shares commonalities and differences, which can be learned jointly.

\paragraph{Multi-task deep neural network (MT-DNN)}  A deep neural network has a wider and deeper architecture as compared to the traditional artificial neural network --  it consists of more layers and more neurons in each layer and reveals the facets of input features at different levels. As for this multitask framework, different tasks share the first few dense layers, on top of which individual predictor is attached to for each specific task (one for partition coefficient and one for aqueous solubility). An illustration of the multitask deep neural network is shown in Fig. \ref{fig:MTDNN_arch}.

In this study, we have a total of 2 tasks -  one for logP prediction and the other for logS prediction. Suppose that there are $N_t$ molecules in $t$-th task and the $i$-th molecule for $t$-th task can be represented by a topological feature vector $\mathbf{x}_i^t$. Given the training data $(\mathbf{x}_i^t, y_i^t)_{i=1}^{N_t}$, where $t={1,2}$, $i={1,...,N_t}$, with $\mathbf{y}_i^t$ being experimental value (logP or logS) for the $i$-th molecule of task $t$,  the objective topological learning is to minimize the function for different tasks: 
\begin{equation}
\sum_{i=1}^{N_t} L(y_i^t, f^t(\mathbf{x}_i^t; \{\mathbf{W}^t,\mathbf{b}^t\} )) 
\end{equation}
where $f^t$ is a functional of topological feature vectors to be learned, parametrized by weight matrix $\mathbf{W}^t$ and bias term $\mathbf{b}^t$, and $L$ can be cross entropy loss for classification and mean squared error for regression. Since logP and logS are measured quantitatively,  the loss function ($t=1$  or $2$) to be minimized is defined as: 
\begin{align}\label{loss}
\mathrm{Loss \hspace{.03in} of \hspace{.03in}}  \mathrm{Task\hspace{.03in} }t 
 = \frac{1}{2}\sum_{i=1}^{N_t} (y_i^t - f(\mathbf{x}_i^t; \{\mathbf{W}^t,\mathbf{b}^t\} )^2 
\end{align}
A wide range of parameters for MT-DNN has been tested and following parameters in Table \ref{tab:params} are used for training and testing in this study although our parameter search is by no means exhaustive:

\begin{table}[!ht]
\centering
\caption{Parameters for MT-DNN.}
\begin{tabular}{|c|c|}
\hline
\# of hidden layers & 7 \\ \hline
\# of neurons on each dense layer & 1000 for first 4 layers and 100 for the next 3 layers \\  \hline
\# of neurons on output layer & 2 with linear activation  \\ \hline
Learning rate & 0.0001 \\
\hline 
\end{tabular}
\label{tab:params}
\end{table}

Note that Adam (adaptive momentum estimation) \cite{kingma:2014} is used as gradient update method. For stability purpose during training, all data were normalized with zero-mean and unit variance.

\paragraph{Remark} Although the units for these two tasks are different (partition coefficient uses log unit as units, and solubility prediction use $\log_{10} \rm{mol/L}$ as units), we can still train a MT-DNN model for them simultaneously - shared layers learn their inherent commonalities while individual layers deal with their differences.

\subsection{Evaluation metrics}
Several different evaluation metrics, including root mean squared error (RMSE), Pearson correlation coefficient (R) and mean unsigned error (MUE), were used to evaluate the performances of different models. These methods are defined as below:
\begin{equation}
\label{eq:R2}
R = \frac{\sum_{i=1}^N(\log X_i^{\rm Pred} - \overline{\log X^{\rm Pred}})(\log X_i^{\rm Expl} - \overline{\log X_i^{\rm Expl}})}{\sqrt{\sum_{i=1}^N(\log X_i^{\rm Pred} - \overline{\log X^{\rm Pred}})^2}\sqrt{\sum_{i=1}^N(\log X_i^{\rm Expl} - \overline{\log X^{\rm Expl}})^2}},
\end{equation}
\begin{equation}
\label{eq:RMSE}
{\rm RMSE}=\sqrt{\frac{1}{N}{\sum_{i=1}^N \left(\log X_i^{\rm Pred}-\log X^{\rm Expl}_i\right)^2}}
\end{equation}
and
\begin{equation}
\label{eq:MUE}
{\rm MUE}=\frac{1}{N}{\sum_{i=1}^N \left|\log X_i^{\rm Pred}-\log X^{\rm Expl}_i\right|},
\end{equation} 
where $\log X$ represents $\log P$ or $\log S$, $N$ is the total number of molecules in the test set, $\log X^{\rm Expl}_i$ and $\log X_i^{\rm Pred}$ stand for the experimental and predicted value for the $i$th molecule, respectively, $\overline{\log X^{\rm Pred}}$ and $\overline{\log X^{\rm Expl}}$ is the average of predicted and experimental value for the entire test set, respectively.  
  
Additionally, Tetko {\it et al.} \cite{mannhold:2009} proposed and an additional metric based on the difference between experimental and predicted $\log P$ ($\Delta \log P$) was proposed. The percentage within each error range was considered. 
\begin{itemize}
\item If $|\Delta \log P| < 0.5$, prediction is considered to be ``acceptable"; 
\item If $0.5 \le |\Delta \log P| < 1.0$, prediction is considered to be ``disputable";
\item If $|\Delta \log P| \ge 1.0$, prediction is considered to be ``unacceptable".
\end{itemize} 
As a result, we use this metric along with $R^2$ to evaluate model performances on Star set and Non-star set described below.

\section{Results} \label{sec:results}

In this section, we present the results of the proposed ESPH methods in conjugation with random forest and  multi-task deep neural networks for variety of data sets, including partition coefficient and solubility test sets. Otherwise stated, different tasks are trained together in the same network. Besides, we would like to introduce some notations for easier reference. ESTD-1 contains 61 Betti-0 bar-based ESTDs (Group 1 in Table \ref{tab:features}), ESTD-2 contains all ESTDs listed in Table \ref{tab:features} (Group 1-3).

\subsection{Partition coefficient prediction} 

\paragraph{Training set cross-validation} In order to have an idea of how our topological representation would work for partition coefficient, a 10-fold cross-validation is performed using baseline method GBDT. Note that 50 runs were done to achieve the final results  as randomness is involved and the results are summarized in Table \ref{tab:crossval}.
\begin{table}[!ht]
\centering
\begin{tabular}{|c|c|c|c|}
\hline
Method  & $R^2$  & RMSE & MUE \\ \hline
{\bf GBDT-ESTD} & 0.924 & 0.45 & 0.32 \\
XLOGP3-AA \cite{chengtj:2007} & 0.904 & 0.50 & 0.39 \\ \hline
\end{tabular}
\caption{Results of 10-fold cross validation on the partition coefficient training set, $N=8199$}
\label{tab:crossval}
\end{table}
It can be seen that our descriptors better than XLOGP3 software \cite{chengtj:2007} given the same training data, and thus demonstrates great predictive power. It would be very interesting to see the performances of our MT-DNN compared to XLOGP3 and GBDT. 

\paragraph{FDA set} The first test set that we would like to apply our model to is the FDA test set. A molecule that contains Hg was dropped due to the difficulty of computation. A major challenge of this set is that its structures are more complex than that of the training set, and the partition coefficient range spans over nearly 12 units. A series of prediction methods \cite{chengtj:2007} including our multi-task neural networks, is applied to this set and their results are summarized in Table \ref{tab:fda} for a comparison with ours. 

\begin{table}[!ht]
\centering
\begin{tabular}{|c|c|c|c|}
\hline
 Method  & $R^2$  & RMSE & MUE \\ \hline
{\bf MT-ESTD-1} & {\bf 0.920} & {\bf 0.57} & {\bf 0.28} \\
{\bf MT-ESTD$^{+}$-1} & {\bf 0.909} & {\bf 0.60} & {\bf 0.27} \\
{\bf MT-ESTD$^{+}$-2} & {\bf 0.909} & {\bf 0.60} & {\bf 0.34} \\
ALOGPS  & 0.908  & 0.60 & 0.42 \\
{\bf MT-ESTD-2} & {\bf 0.891} & {\bf 0.66} & {\bf 0.44} \\
XLOGP3  & 0.872  & 0.72 & 0.51 \\  
XLOGP3-AA &  0.847 & 0.80 & 0.57 \\
CLOGP & 0.838 & 0.88 & 0.51 \\
TOPKAT & 0.815 & 0.88 & 0.56 \\
ALOGP98 & 0.80 & 0.90 & 0.64 \\
KowWIN & 0.771 & 1.10 & 0.63 \\
HINT & 0.491 & 1.93 & 1.30 \\ \hline
\end{tabular}
\caption{Results of different logP prediction methods on 406 FDA-approved drugs \cite{chengtj:2007}, ranked by $R^2$. Two molecules were dropped for our model evaluation due to feature generation failure of ChemoPy}
\label{tab:fda}
\end{table}

As we can see from Table \ref{tab:fda}, our multi-task model gives the best prediction in terms of R${^2}$, RMSE and MUE. Specifically the small MUE of our model (0.29 log units) indicates that our predictions are less biased than other methods tested, except for some outliers. Also note that the training set is completely independent of the test set which shows the applicability of our multi-task architecture. We also build models with the same architecture when additional molecules gathered from NIH-database are included as there is no guarantee that ALOGPS are completely independent of the testset. It turns out that the accuracy can be greatly improved. 

\paragraph{Star set and Non-star set} Star set and Non-star set were proposed by Tetko \cite{mannhold:2009} as two benchmark sets for evaluating partition coefficient models. Over 20 different models were tested on these two sets. It should be emphasized that for these sets, different models are trained on different training sets and their overlap with the test sets is unknown. Thus it makes more sense to merge our 8199 training set with additional molecules in NIH database and see how additional training data can benefit the overall performances. Results of different models on these two sets can be found in Table \ref{tab:overall_result}. Notice that models trained with additional data are labeled with superscript *. 
  
For star set, we achieve RMSE of 0.53 log units with other popular commercial software packages such as ACD/logP and CLOGP, in addition to a high acceptable prediction percentage (75\%, rank 3) as well as a low unacceptable rate (5 \%, rank 3). For non-star set, most methods do not give accurate predictions as the structures in this set are relatively new and complex. Our 49\% acceptable rate ranks number 2 among all predictors, though RMSE is relatively high due to a few large outliers (rank 5). The results are satisfactory as commercial software packages generally use a much larger training set than that ours. In general, when there exist more overlapped molecules in the training set, the test results will  be significantly improved. As Table \ref{tab:overall_result} indicates, our MT-DNN-ESTD$^+$ models achieve a substantial improvement over XLOP3 for star set, while maintaining the same level of accuracy for non-star set. The performances of our MT-DNN-ESTD$^+$ models suggest that our models are able to predict $\log P$ and $\log S$ accurately. 
In fact,  the predictive power can potentially be further improved once more molecules are incorporated into the training set.  Thus we also extend our original 8199 training set by adding molecules in both Star set and Non-star set. When we use the extended training set, much higher accuracy is achieved as listed in Table \ref{tab:overall_result}. Again we observe that  MT-DNN outperforms RF with the same set of features. 

\begin{table}[h!]
\centering
\begin{tabular}{C{4cm}C{2.0cm}C{1cm}C{1cm}C{1cm}@{\extracolsep{4pt}}@{} C{2.0cm}C{1cm}C{1cm}C{1cm}} %{m{2cm} m{2cm} m{2cm} m{1.5cm} r}
% \hline
& \multicolumn{4}{c}{Star Set ($N=223$)} & \multicolumn{4}{c}{Non-Star Set ($N=43$)} \\
\cline{2-5} \cline{6-9}
& & \multicolumn{3}{c}{\multirow{1}{*}{\% of Molecules}} & & \multicolumn{3}{c}{\multirow{1}{*}{\% of Molecules}}\\ 
& & \multicolumn{3}{c}{Within Error Range} &  & \multicolumn{3}{c}{Within Error Range} \\
\cline{3-5} \cline{7-9}
Method & RMSE & < 0.5 & <1 & > 1 & RMSE & < 0.5 & <1 & > 1\\ \hline
AB/LogP & 0.41 & 84 & 12 & 4 & 1.00 & 42  & 23 & 35\\
S+logP & 0.45 & 76 & 22 & 3 & 0.87 & 40 &35  & 26\\
{\bf MT-ESTD$^+$-1-AD} & 0.49 & 77 &  16 & 7 & 0.98 & 49 & 19 & 33 \\
{\bf MT-ESTD$^+$-2} & 0.49 & 74 &  21 & 5 & 0.97 & 49 & 23 & 28 \\
ACD/logP & 0.50 & 75 & 17 & 7 & 1.00  & 44  &32 &23\\
CLOGP & 0.52 & 74 & 20 & 6 & 0.91 & 47 & 28  & 26\\
VLOGP OPS & 0.52 & 64 & 21 & 7 & 1.07 & 33 & 28 & 26\\
ALOGPS  & 0.53  & 71 & 23 & 6 & 0.82 & 42  &30 & 28\\
{\bf MT-ESTD$^+$-1} & 0.53 & 75 &  17 & 8 & 0.97 & 47 & 28 & 26 \\
{\bf MT-ESTD-1-AD} & 0.53 & 73 &  18 & 9 & 1.00 & 37 & 30 & 33 \\
{\bf MT-ESTD-2-AD} & 0.53 & 71 &  19 & 9 & 1.01 & 47 & 19 & 35 \\
{\bf MT-ESTD-1} & 0.55 & 72 &  18 & 10 & 1.01 & 33 & 28 & 40 \\
{\bf MT-ESTD-2} & 0.56 & 66 &  23 & 11 & 1.06 & 35 & 33 & 33 \\
MiLogP & 0.57 & 69 & 22 & 9 & 0.86 & 49 & 30 & 21\\
XLOGP3  & 0.62 & 60 & 30 & 10 &  0.89 &47  & 23 & 30 \\
KowWIN & 0.64 & 68 &  21 & 11 & 1.05  & 40  & 30  & 30 \\
CSLogP & 0.65 & 66 & 22 & 12 & 0.93 & 58 & 19  & 23\\
ALOGP & 0.69 & 60 & 25 & 16 & 0.92 & 28 & 40  & 33\\
MolLogP & 0.69 & 61 & 25 & 14 & 0.93 & 40 & 25  & 26\\
ALOGP98 & 0.70 & 61 & 26 & 13 & 1.00 & 30 & 37  & 33 \\
OsirisP & 0.71 & 59 & 26 & 16 & 0.94 & 42 & 26 & 33\\
VLOGP & 0.72 & 65 & 22 & 14 & 1.13 & 40 & 28 & 33\\
TLOGP & 0.74 & 67 & 16 & 13 & 1.12 & 30 & 37 & 30\\
ABSOLV & 0.75 & 53 & 30 & 17 & 1.02  & 49 & 28 & 23\\
QikProp& 0.77 & 53 & 30 & 17 & 1.24 & 40 & 26  & 35\\
QuantlogP & 0.80 & 47 & 30 & 22 & 1.17 & 35 & 26 & 40\\
SLIPPER-2002 & 0.80 & 62 & 22 & 15 & 1.16 & 35& 23& 42\\
COSMOFrag & 0.84 & 48 & 26 & 19 & 1.23 & 26 & 40 & 23\\
XLOGP2 & 0.87 & 57 & 22 & 20 & 1.16 & 35 & 23  & 42\\
QLOGP & 0.96 & 48 & 26 & 25 & 1.42 & 21 & 26 & 53\\
VEGA & 1.04 & 47 & 27 & 26 & 1.24 &  28 & 30 & 42\\
CLIP & 1.05 & 41 & 25  & 30  & 1.54 & 33 & 9 & 49\\
LSER & 1.07 & 44 & 26 & 30 & 1.26 & 35 & 16 & 49\\
MLOGP(Sim+) & 1.26  & 38 & 30 & 33 & 1.56 & 26 & 28 & 47\\
NC+NHET & 1.35 & 29 & 26 & 45 &  1.71 & 19 & 16 & 65\\
SPARC & 1.36 & 45 & 22 & 32 &1.70 & 28 & 21 & 49\\
%MLOGP(Dragon) & 1.52 &39 & 26 & 35 & 2.45 & 23& 30  & 47\\
HINTLOGP & 1.80 & 34 & 22 & 44 & 2.72 & 30 & 5 & 65 \\ \hline
\end{tabular}
\caption{Benchmark test results \cite{mannhold:2009} of different logP prediction methods on both Star set molecules and Non-Star set molecules. The superscript $^*$ means that Star molecules and Non-star molecules were included in the training.}
\label{tab:overall_result}
\end{table} 

\subsection{Aqueous solubility prediction}
To evaluate the performances of solubility our models, several datasets are used, derived from Wang {\it et al.} \cite{wangjm:2007} and Hou {\it et al.} \cite{hou:2004}. For leave-one-out validation, only baseline method is used. For 10-fold cross-validation, the 9 remaining folds are trained together with the partition coefficient training set when evaluating the remaining fold with MT-DNN architecture. 

\subsubsection{1708 set in Ref. \cite{wangjm:2007}} For this dataset, both leave-one out and 10-fold cross validations are carried out in order to evaluate the performance of our models. 

\paragraph{Leave-one-out} As MT-DNN requires a lot of computational resources, only baseline method GBDT is used for leave-one-out prediction. We use 4000 trees and 0.10 learning rate as training parameters to develop models and following results in Table \ref{tab:sol1} are achieved.

\begin{table}[!ht]
\centering
\begin{tabular}{|c|c|c|c|}
\hline
Method  & $R^2$  & RMSE & MUE \\ \hline
{\bf GBDT$^+$-1-AD} & 0.931 & 0.543 & 0.389 \\ \hline
{\bf GBDT$^+$-2-AD} & 0.929 & 0.551 & 0.389 \\ \hline
{\bf GBDT$^+$-2} & 0.910 & 0.621 & 0.457 \\ \hline
ASMS-LOGP\cite{wangjm:2007} & 0.897 & 0.664 & 0.505 \\ \hline
{\bf GBDT$^+$-1} & 0.893 & 0.683 & 0.494 \\ \hline
ASMS\cite{wangjm:2007} & 0.884 & 0.707 & 0.547 \\ \hline
\end{tabular}
\caption{Leave-one-out test on the 1708 solubility dataset}
\label{tab:sol1}
\end{table}

\paragraph{10-fold cross-validation} As MT-DNN and baseline method GBDT involves randomness, we run MT-DNN and GBDT 50 times and report mean performances for all metrics. The results are summarized in Table \ref{tab:sol2}. It is observed that our models yield more accurate and robust predictions than ASMS and ASMS-LOGP models do, improving the $R^2$ from 0.884 to 0.925. Additionally we also notice that there generally exists an improvement of MT-ESTD models over GBDT models, though, not as significant as what we see in previous partition coefficient prediction.

\begin{table}[!ht]
\centering
\begin{tabular}{|c|c|c|c|}
\hline
 Method  & Mean $R^2$ (RMSD)  & Mean RMSE (RMSD) & Mean MUE (RMSD)\\ \hline
{\bf MT-ESTD$^+$-1} & 0.925 (0.001) & 0.568 (0.005) & 0.393 (0.003) \\ \hline
{\bf MT-ESTD$^+$-2} & 0.924 (0.003) & 0.571 (0.010) & 0.395 (0.004) \\ \hline
{\bf GBDT$^+$-1} & 0.924 (0.002) & 0.572 (0.006) & 0.408 (0.005) \\ \hline
{\bf GBDT$^+$-2} & 0.923 (0.002) & 0.571 (0.006) & 0.408 (0.005) \\ \hline
{\bf MT-ESTD-1} & 0.908 (0.002) & 0.630 (0.005) & 0.466 (0.003) \\ \hline
{\bf GBDT-2} & 0.904 (0.002) & 0.642 (0.008) & 0.469 (0.005) \\ \hline
{\bf MT-ESTD-2} & 0.902 (0.002) & 0.649 (0.007) & 0.466 (0.005) \\ \hline
{\bf GBDT-1}  & 0.889 (0.003) & 0.697 (0.009) & 0.502 (0.005) \\ \hline
ASMS\cite{wangjm:2007} & 0.884 (0.021) & 0.699 (0.054) & 0.527 (0.034) \\ \hline
ASMS-LOGP \cite{wangjm:2007} & 0.869 (0.022) & 0.742 (0.053) & 0.570 (0.034) \\ \hline
\end{tabular}
\caption{10-fold cross-validation on the 1708 solubility dataset}
\label{tab:sol2}
\end{table}

\subsubsection{Data set in Ref. \cite{hou:2004}}
We test our models on data set proposed by Hou {\it et al.} \cite{hou:2004}, where training and test sets were predefined to cover a variety of molecules. Test set 1 contains 21 commonly used compounds of pharmaceutical and environmental interest \cite{klopman:1992} and is to be trained on the original 1290 molecules. Test set 2 contains 120 molecules that were used to develop Klopman and Zhu's group contribution model \cite{klopman:2001}. As Hou  {\it et al.} \cite{hou:2004} suggested, we remove 83 molecules that overlap with test set 2 from the training set to make prediction independent and unbiased. This reduces the size of training set for test set 2 to 1207.

\paragraph{Test set 1} Table \ref{tab:sol3} shows the performances of different models on test set 1. Our MT-ESTD models perform similarly to Drug-LOGS method, while achieving improvement over Klopman and Zhu's MLR method \cite{klopman:2001} with ESTDs. 
\begin{table}[!ht]
\centering
\begin{tabular}{|c|c|c|}
\hline
Method  & R &  RMSE  \\ \hline
{\bf MT-ESTD$^+$-1} & 0.94 & 0.69 \\ \hline 
Drug-LOGS\cite{hou:2004} & 0.94 & 0.64\\ \hline
{\bf GBDT$^+$-2} & 0.94 & 0.71 \\ \hline 
{\bf MT-ESTD$^+$-2} & 0.93 & 0.75 \\ \hline 
{\bf GBDT$^+$-1} & 0.93 & 0.76 \\ \hline 
{\bf MT-ESTD-2} & 0.92 & 0.79 \\ \hline 
Klopman MLR \cite{klopman:2001} & 0.92 & 0.86 \\ \hline 
{\bf GBDT-2} & 0.92 & 0.85 \\ \hline 
{\bf MT-ESTD-1} & 0.91 & 0.82 \\ \hline 
{\bf GBDT-1} & 0.84 & 1.07 \\ \hline 
\end{tabular}
\caption{Results of test set 1 \cite{hou:2004}, where MUE was not reported}
\label{tab:sol3}
\end{table}

\paragraph{Test set 2} The results of test set 2 are summarized in Table \ref{tab:sol4}. For this dataset, our MT-ESTD models gives satisfactory results with a high Pearson correlation over 0.97 across all ESTD combinations. Such results indicate that our methods are   applicable to a wide variety of molecules.

\begin{table}[!ht]
\centering
\begin{tabular}{|c|c|c|c|}
\hline
Method  & R &  RMSE &  MUE \\ \hline
{\bf MT-ESTD$^+$-1} & 0.97 & 0.65 & 0.47 \\ \hline 
{\bf MT-ESTD$^+$-2} & 0.97 & 0.67 & 0.48 \\ \hline 
{\bf MT-ESTD-1} & 0.97 & 0.70 & 0.50 \\ \hline 
{\bf MT-ESTD-2} & 0.97 & 0.71 & 0.53 \\ \hline 
{\bf GBDT$^+$-2} & 0.97 & 0.73 & 0.50 \\ \hline
{\bf GBDT$^+$-1} & 0.96 & 0.76 & 0.52 \\ \hline 
Drug-LOGS \cite{hou:2004} & 0.96 & 0.79 & 0.57 \\ \hline
{\bf GBDT-2} & 0.96 & 0.79 & 0.60 \\ \hline 
{\bf GBDT-1} & 0.96 & 0.82 & 0.58 \\ \hline 
Group contribution \cite{klopman:2001} & 0.96 & 0.84 & 0.70 \\ \hline 
\end{tabular}
\caption{Results of Klopman and Zhu's test set 2}
\label{tab:sol4}
\end{table}

\section{Discussion} \label{sec:discussion}
In this section, we discuss the outcome of trained models from several aspects. The emphasis is put on the predictive power of our ESTDs in terms of MT-DNN and how topology based multitask learning can improve partition coefficient and aqueous solubility predictions. 

%\paragraph{Feature importance analysis}  One major concern for quantitative learning is feature importance analysis. It is of significance to determine if a set of features is sufficient for modeling a target value or further removing redundant features can help improve accuracy. To this concern, we choose 41 different sets of features, ranked by their feature importance calculated by our baseline methods, and train separate models to see their predictive performances on test sets. 

\subsection{ESTDs for small molecules} As the previous results indicated, there exists a common feature representation for both partition coefficient and solubility prediction. Our features come from two different categories -- one that is computed solely by ESPH, and the other one that has been widely used in the development of QSAR models. Although the number of ESTDs (121) is small, it turns out that our topological feature representation of molecules has a very strong predictive power as compared to baseline method. Our ESTDs highlight atom type information and keep track of the formation of chemical bond as well as ring structures of a given molecule. Indeed,   by constructing topological space of a molecule, ESTDs are able to extract useful features from  ESPH computation.
In fact, more detailed topological representations which more precisely reflect the covalent bond length, hydrogen bond strength  and van der Waals interaction can constructed via a small bin size \cite{ZXCang:2017c}.   

\subsection{Multitask learning} The goal of multitask learning is to learn commonalities between different tasks, and to simultaneously improve model performances. In this study, partition coefficient and aqueous solubility were trained jointly and substantial improvements over singletask models are observed. Our results suggests that there exists shared information across these two tasks that can benefit prediction accuracy. Indeed, the original motivation for predicting logP and logS is that both coefficients closely relate to the extent to which a compound dissolves in solvent. By comparing our MT-DNN with gradient boosting tree, we find that it is beneficial to learn partition coefficient and aqueous solubility models together. Our MT-ESTD models achieve satisfactory results on various partition coefficient and aqueous solubility data sets, some of which are the state-of-the-art to our knowledge. Our ESTDs alone can give very accurate predictions, bringing us new insights by element specific persistent homology computations. In addition to ESTDs, commonly-used 2D descriptors also help to improve the overall accuracy. The attempt to learn these two related measurements together gives a boost over learning them separately and the improvement is validated on most data sets. 

\section{Conclusion}
Partition coefficient and aqueous solubility are among most important physical properties of small molecules and have significant applications to drug design and discovery in terms of lipophilic efficiency. Based on chemical and physical models, a wide variety of computational methods has been developed in the literature for the theoretical predictions of partition coefficient and aqueous solubility.  

Present work introduces an algebraic topology based method, element specific persistent homology (ESPH), for simultaneous  partition coefficient and aqueous solubility predictions. ESPH  offers an unconventional representation of small molecules in terms of multiscale and multicomponent topological invariants.  Here the multiscale representation is inherited from persistent homology, while the multicomponent formulation is developed to retain essential chemical information during the topological simplification of molecular geometric complexity. Therefore, the present ESPH gives a unique representation of small molecules that cannot be obtained by any other method. Although ESPH representation of molecules cannot by  literally translated into a physical interpretation, it systematically and comprehensively encipher chemical and physical information of molecules into scalable topological invariants, and thus is  ideally suited for machine learning/deep learning algorithms to decipher such information.  
 
To predict partition coefficient and aqueous solubility, we integrate ESPH with advanced machine learning methods, including gradient boosting tree, random forest,   and deep neural networks to construct topological learning strategies. Since partition coefficient and aqueous solubility are highly correlated to each other, we develop a common set of ESPH descriptors, called element specific topological descriptors  (ESTDs), to represent both properties. This approach enables us to perform simultaneous predictions of partition coefficient and aqueous solubility using a topology based  multi-task deep learning strategy.

To test the representational of ESPH and the predictive power of the proposed topological multi-task deep learning strategy, we consider some commonly used data sets, including two benchmark test sets, for partition coefficient,  as well as additional solubility data sets. Extensive cross validations and benchmark tests indicate that the proposed topological multi-task strategy offers some of the most accurate predictions of partition coefficient and aqueous solubility. 

\vspace{0.5cm}
\section*{Availability}
Our software is available as an online server at {\url{http://weilab.math.msu.edu/TopP-S/}. 

\section*{Acknowledgment}

This work was supported in part by NSF Grants DMS-1721024 and  IIS-1302285,  and MSU Center for Mathematical Molecular Biosciences Initiative.

\vspace{1cm}
\bibliographystyle{ieeetr}
\bibliography{refs}

\end{document}